\documentclass{article}

\usepackage{PRIMEarxiv}

\usepackage[utf8]{inputenc} 
\usepackage[T1]{fontenc}    
\usepackage{hyperref}       
\usepackage{url}            
\usepackage{booktabs}       
\usepackage{amsfonts}       
\usepackage{nicefrac}       
\usepackage{microtype}      
\usepackage{lipsum}
\usepackage{fancyhdr}       
\usepackage{graphicx}       
\usepackage{algpseudocode,algorithm,algorithmicx}

\graphicspath{{media/}}     

\title{Inflexible Multi-Asset Hedging of incomplete market 

}

\author{
  Ruochen Xiao\textsuperscript{1}, Qiaochu Feng\textsuperscript{1} \\
  Shanghai University \\
  Shanghai\\
  \texttt{\{Ruochen Xiao\}ninaxiao1208@shu.edu.cn} \\
   \And
  Ruxin Deng\textsuperscript{3} \\
  University of Chinese Academy of Sciences \\
  Beijing\\
}

\begin{document}
\maketitle

\begin{abstract}
Models trained under assumptions in the complete market usually don't take effect in the incomplete market. This paper solves the hedging problem in incomplete market with three sources of incompleteness: risk factor, illiquidity, and discrete transaction dates. A new jump-diffusion model is proposed to describe stochastic asset prices. Three neutral networks, including RNN, LSTM, Mogrifier-LSTM are used to attain hedging strategies with MSE Loss and Huber Loss implemented and compared.As a result, Mogrifier-LSTM is the fastest model with the best results under MSE and Huber Loss.
\end{abstract}


\section{Introduction}
Trading in real market has lots of risks and limits such as transactions costs, discrete time hedging dates, illiquidity and non-tradable risk factors. These factors make results under the completeness assumption unreliable in most of time. This paper aims to solve hedging problems with three sources of incompleteness: volume risks, discrete tradable dates, and illiquidity constraints.

Based on Merton's jump-diffusion model, many studies have been done over the simulation of extreme value movements. In \cite{2008Bilateral},bilateral gamma distribution have excellent degree of fitting the German stock index(DAX).In this paper, a degraded bilateral gamma distribution: variance gamma distribution \cite{2001Purely} is taken to simulate the jump size in the classic jump-diffusion model.

Previous studies have made great achievements in establishing multi-dimension stock price series models(Fecamp et al,2020\cite{fecamp2020deep}) and fitness of one-dimension model into real financial market has also been proved(Gao,2021\cite{0Inflexible}).However, the adaptation of the multi-dimension model has not been carried out with the existence of off-diagonal covariance matrix which describes the inter-asset Brownian motion.In this paper, we wish to solve this problem by using neural network. 

Neural networks have made significant achievements in solving non-linear PDE problems. We use three neural networks: RNN,LSTM,Mogrifier-LSTM to deal with the hedging problem in one dimension. Considering the cost of time and disk usage, we expand the one-dimension problem to multi-dimension one applying the best-performed algorithm: Mogrifier-LSTM. Despite the fact that the loss function of the algorithm converges within hundreds of iteration, the computer capability remains a limitation to the application of the Mogrifier-LSTM.

\section{Jump-Diffusion Model}
\label{sec:model}
\subsection{Derivation of Jump-Diffusion Model}
\label{subsec:JD}
A financial market with continuous operating time is set up in the time horizon $0<t<T$ and the probability space $(\Omega,\mathcal{F},\mathcal{P})$ where Ft denotes the available information at time t. In this market, a risk-free asset (the bond) and d risky assets (the stock) is considered. The risk-free rate $ r (r>0)$ is defined under the risk neutral measure while yield and transaction cost is not included. 

To quantify the risky asset, we make modifications on Merton's jump-diffusion\cite{Matsuda2004IntroductionTM} model to describe risky assets . The asset price at time t is denoted by $S_t$($0<t<T$) in the matrix form:

\begin{equation}
    dS_t=diag(S_{t-})\mu dt + diag(S_{t-})\theta dW_t +diag(S_{t-})dJ_t
\end{equation}
\begin{equation}
    J_t=\sum^{N_t}_{i=1}exp(U_i)-1
\end{equation}
and the integral form:
\begin{equation}
    S_t=S_0+ \int^t_0 diag(S_{t-})\mu d\tau+\int^t_0 diag(S_{t-})\mu dW_{\tau}+\int^t_0 diag(S_{t-})\mu dJ_{\tau}
\end{equation}
$W_t$ is a standard Brownian motion process. $\mu$ is a $d\times1$ matrix, which denotes the mean matrix of history data of d assets and $\sigma$ is a $d\times d$ matrix which denotes the covariance matrix of d assets.

$J_t=\sum^{N_t}_{i=1}exp(U_i)-1$ is a compound poisson process.There are two sources of randomness in this term. First, the Poisson process $dN_t$ with the parameter $\lambda$ (i.e. average number of jumps per year) allows the asset price to jump randomly. This means that the time internal within random jumps follow an exponential distribution with the same parameter $\lambda$ as the Poisson process. The other one is the random jump size 
$exp(U_i)-1$. Given the importance of heavy tail and negative skewness in the movement of extreme values, a left heavy tail is involved to simulate the considerable probability of extreme loss. Bilateral-gamma processes which are defined as the difference of two independent Gamma process is suggested in \cite{2008Bilateral} as a replacement of the classic Brownian motion. In this paper we use it as a simulation of the jump size. 
\begin{equation}
    (U_i \sim i.i.d.\Gamma(\alpha^{+},\lambda^{+};\alpha^{-},\lambda^{-}))
\end{equation}
where 
$$\alpha^{+}=\alpha^{-}>0 \quad and \quad \lambda^{+},\lambda^{-}>0 $$
When $\alpha^{+}=\alpha^{-}=1$, it becomes double exponential distribution given by Kou(2002)\cite{Kou2002AJM} and the fitness into stock price simulation has been proved in Gao(2021) \cite{0Inflexible}. Therefore ,the case of double exponential distribution could be explained by bilateral-gamma distribution in generalized form. In order to give an explicit density function,  we set $\alpha^{+}$=$\alpha^{-}$=$\alpha$=2 so that the bilateral-gamma distribution degrades into a variance-gamma distribution \cite{2001Purely} comprising three parameters$(\alpha,\lambda_{+},\lambda_{-})$ which enable us to take control over the skewness and kurtosis. The density function of $Y_i(=U_i)$ follows:

\begin{equation}
    f_Y(y)=\frac{{(\lambda^{+} \lambda^{-})}^\alpha}{{(\lambda^{+}+\lambda^{-})}^\alpha {\Gamma(\alpha)}^2} e^{-\lambda^{+} (y-p)} \int_0^{+\infty} v^{\alpha-1}(y-p+\frac{v}{\lambda^{+}+\lambda^{-}})^{\alpha-1}e^{-v} dv \mathbf{1}_{(y>p)}
\end{equation}
\begin{equation}
    +\frac{{(\lambda^{+} \lambda^{-})}^\alpha}{{(\lambda^{+}+\lambda^{-})}^\alpha {\Gamma(\alpha)}^2} e^{\lambda^{-} (y-p)} \int_0^{+\infty} v^{\alpha-1}(y-p+\frac{v}{\lambda^{+}+\lambda^{-}})^{\alpha-1}e^{-v} dv \mathbf{1}_{(y\leq p)}
\end{equation}
where each part of the sum above denotes the density of jump sizes compared with $y=p$.  The four characteristic indicators of this distribution are as follows:
\begin{itemize}
    \item[*] The expectation $$ E(Y) = \frac{\alpha}{\lambda^{+}}-\frac{\alpha}{\lambda^{-}}<\infty$$
    \item[*] The variance $$Var(Y)= \frac{\alpha}{{(\lambda^{+})}^2}+\frac{\alpha}{{(\lambda^{-})}^2}$$
    \item[*] The Charliers skewness $$Skew(Y)=\frac{2(\frac{\alpha}{{(\lambda^{+})}^3}-\frac{\alpha}{{(\lambda^{-})}^3})}{{Var(Y)}^{3/2}}$$
    \item[*] The kurtosis $$ Kur(Y)=3+\frac{6(\frac{\alpha}{{(\lambda^{+})}^4}+\frac{\alpha}{{(\lambda^{-})}^4})}{{Var(Y)}^2}>3$$
\end{itemize}
It can be concluded that variance Gamma distributions are strictly leptokurtic with all parameters positive. $\lambda^+ > \lambda^- >0$ is stated for the case of negative skewness. Also, $E(U)<\infty$ is ensured.

As Merton assumed, the Brownian Motion $W_t$ and the two sources of randomness proposed above are independent of each other. In this model, $\mu$ and $\sigma$ are defined as constant matrix. Since we have d risky assets available for trade,$W_t$  and $N_t$ are both d-dimensional. The Ito’s form of one-dimension asset price in equation (1) is 
\begin{equation}
    S_t=S_0 exp((\mu-\frac{\sigma^2}{2})t+\sigma W_t+\sum_{i=1}^{N_t}U_i)\label{equ:5}
\end{equation}

\subsection{Estimation} \label{sec:sampling}

The estimation is done in one dimension in this part. To estimate the parameters, we focus on the exponent part of the asset price. The log return of assets over $\Delta t$ according to equation7 is:
\begin{equation}
   ln(\frac{S_{t+\Delta t}}{S_t})=(\mu-\frac{\sigma^2}{2}) \Delta t+\sigma (W_{t+\Delta t}-W_t)+\sum_{i=N_{t+1}}^{N_{t+\Delta t}}U_i
\end{equation}
Under daily observations, $\Delta_t$ is set to be 1 day ($\frac{1}{252}$ year). To derive the density function, the jump part within a small interval $\Delta_t$ is approximated by a Bernoulli random variable B with $Pr[B=1]=\lambda \Delta_t$ and $Pr[B=0]=1- \lambda \Delta_t$. According to \cite{0Inflexible} the approximation of the daily log return is:
\begin{equation}
    ln(\frac{S_{t+\Delta t}}{S_t})\approx(\mu-\frac{\sigma^2}{2}) \Delta t+\sigma \sqrt{\Delta t} \mathit{Z}+\mathit{BY}\label{equ:7}
\end{equation}
Z is a random variable following the standard normal distribution. The density function of the RHS of 9 is as follows and the derivation of this function is presented in appendix:

\begin{equation}
    f_{RHS}(x)= \frac{1-\lambda \Delta t}{\sigma \sqrt{\Delta t}} \phi(\frac{x-\Delta t(\mu-\frac{\sigma^2}{2})}{\sigma \sqrt{\Delta t}})+ \lambda \Delta t k\times e^{-\lambda^{+}(x-p)} e^{\frac{\Delta t}{2}[2\lambda^{+}(\mu-\frac{\sigma^2}{2})+\lambda^{+} \sigma^2]} \times[(\frac{2}{m}+p)\Phi_1(x-p)+\int_{-\infty}^{x-p}\Phi_1(v)dv]
\end{equation}
\begin{equation}
    +\lambda \Delta t k\times e^{\lambda^{-}(x-p)} e^{-\frac{\Delta t}{2}[2\lambda^{-}(\mu-\frac{\sigma^2}{2})-\lambda^{-} \sigma^2]} \times[(\frac{2}{m}+x)-(\frac{2}{m}+p)\Phi_2(x-p)-\int_{-\infty}^{x-p}\Phi_2(v)dv]\label{density function}
\end{equation}

We did Kolmogorov-Smrinov test in one dimension using data from American stock market since July 2019 to August 2021 to reject the Gaussian distribution. $\lambda$ is derived from the annual outliers from the interval $[-3\sigma, 3\sigma]$. In our data set $\lambda=2$. Applying the Hooke-Jeeves algorithm\cite{2008Hooke}, we maximize the logarithmic likelihood function numerically. The density plot of function 11 and the real empirical data is presented in 1.
Under maximum likelihood estimation(MLE), the parameters are $\mu=0.25$, $\sigma=0.2787$, $\lambda^{+}=0.1834$, $\lambda^{-}=0.1629$, $p=0.0275$.

\begin{figure}
    \centering
    \includegraphics[scale=0.7]{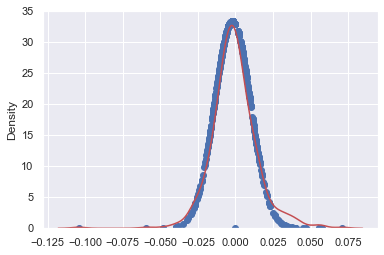}
    \caption{The density plot of log return}
    \label{fig:density plot}
\end{figure}
The red line represents for the density of empirical data and the points stands for the values of our model.

\subsection{Risk-Neutral Measure}\label{subsec:risk neutral}

According to the no-arbitrage principle, a measure change should be made from physical probability space $\mathcal{P}$ to the risk-neutual measure space $\mathcal{Q}$. Jumps are not considered in the Randon-Nikodym Derivative (RND):
\begin{equation}
    \frac{d\mathcal{Q}}{d\mathcal{P}}\arrowvert_t=exp(-\frac{1}{2} \theta^2 t-\theta W_t)
\end{equation}

Also, the measure change should follow:
\begin{equation}
S_s=exp(-r(t-s))\mathbb{E}^{\mathcal{Q}}(S_t|\mathcal{F}_s)
\end{equation}
where $r$ stands for the risk-free rate and $\theta$ is the discount rate. The equations above hold for both one-dimension problem and multi-dimension ones. Note that in multi-dimension problem, $\theta$ is a $d\times 1$ matrix.

Gao(2021)\cite{0Inflexible}  suggested a explicit expression of $\theta$:
\begin{equation}
    \theta\sigma=\mu-r+\lambda(E(Y)-1)
\end{equation}
Then the asset price under risk-neutral measure is:
\begin{equation}
    S_t=S_0 exp((\mu-\frac{\sigma^2}{2})t+\sigma W_t^{\mathcal{Q}}+\sum_{i=1}^{N_t}U_i-\sigma\theta t)
\end{equation}

\subsection{Simulation}
We first did simulations in one dimension on the jump-diffusion models in 2.1 and 2.3 in two years $T=504$ days for 900 times to train neural networks.A part of simulation paths is presented in Fig.2. And a general picture of 500 times simulation is shown in Fig.3.

\begin{figure}
    \centering
    \includegraphics[scale=0.56]{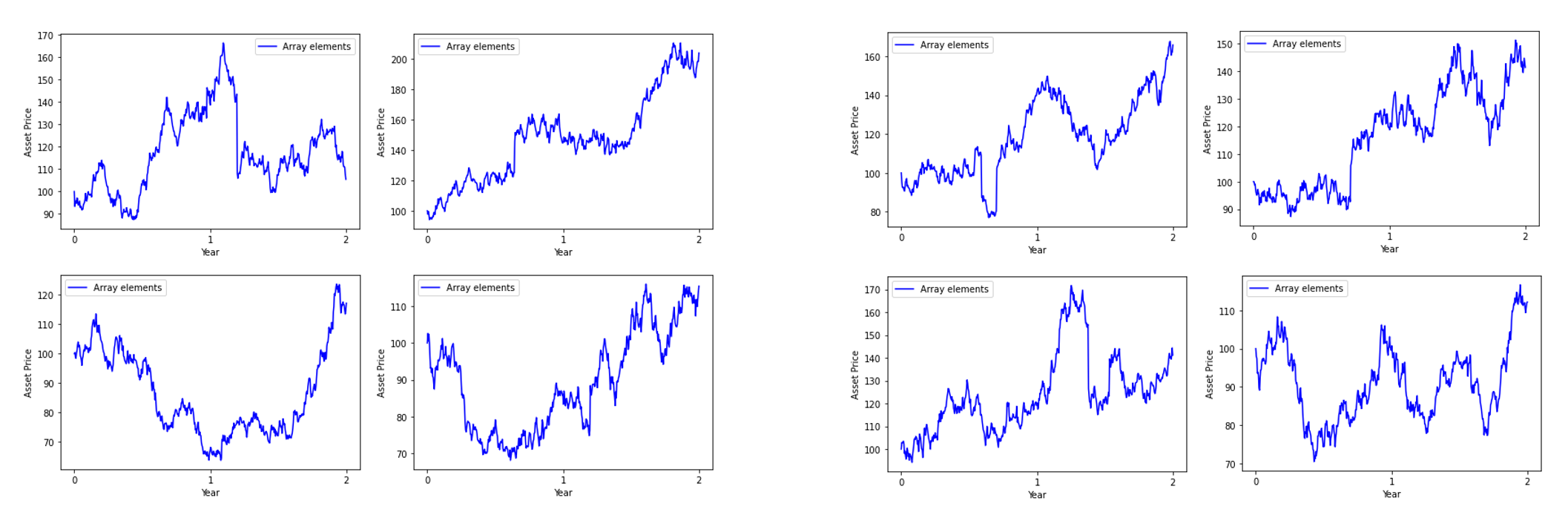}
    \caption{Single Asset price under physical measure(left) and risk-neutral measure(right)}
    \label{fig: sim_one}
\end{figure}

\begin{figure}
    \centering
    \includegraphics[scale=0.75]{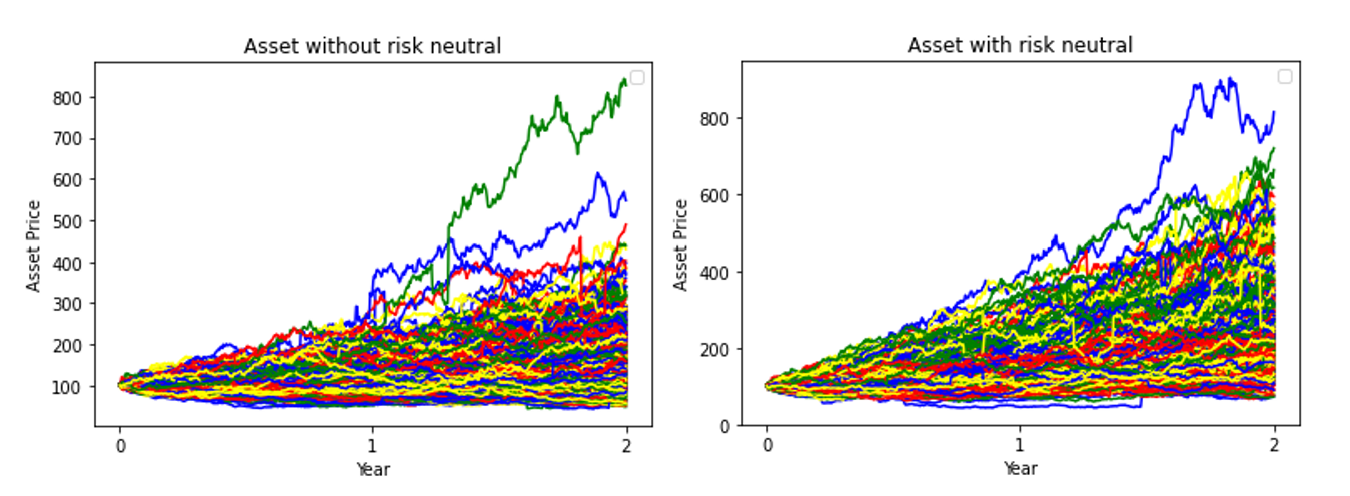}
    \caption{Assets price under physical measure(left) and risk-neutral measure(right)}
    \label{fig: sim_500}
\end{figure}

Fluctuated movements and few steep jumps is observed in both models. In Fig.3, the fluctuation of overall condition is obvious.
In muti-dimension problem, simulation is done for 5 assets. $S_0$ are all set to 100 for the sake of simplicity. The correlation between assets are involved in the covariance matrix $\sigma$ while inter-asset jumps are all considered independent. 5-asset simulations under two models are shown in Fig.4 and Fig.5.

\begin{figure}
    \centering
    \includegraphics[scale=0.6]{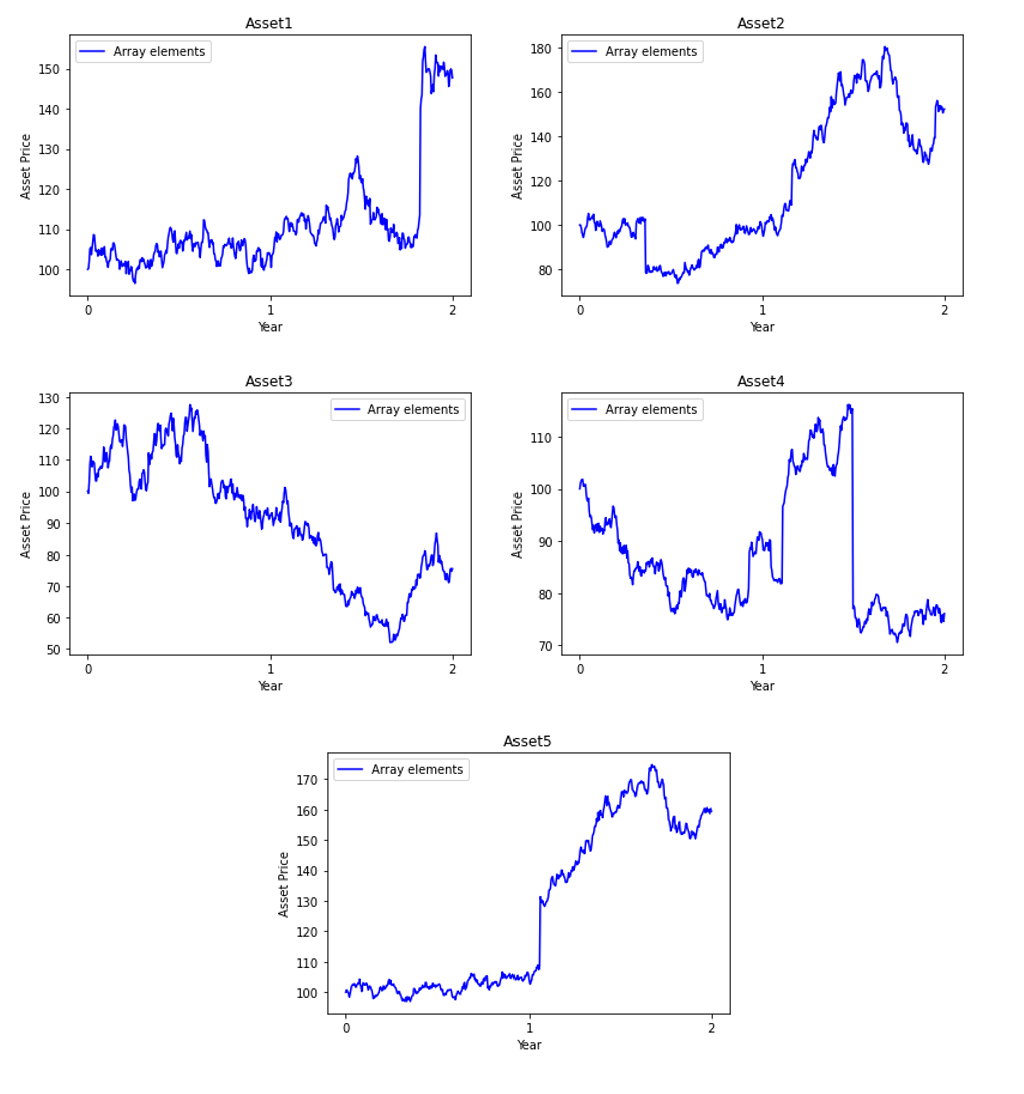}
    \caption{5-asset simulation without risk neutral}
    \label{fig:5assets_wtrn}
\end{figure}

\begin{figure}
    \centering
    \includegraphics[scale=0.6]{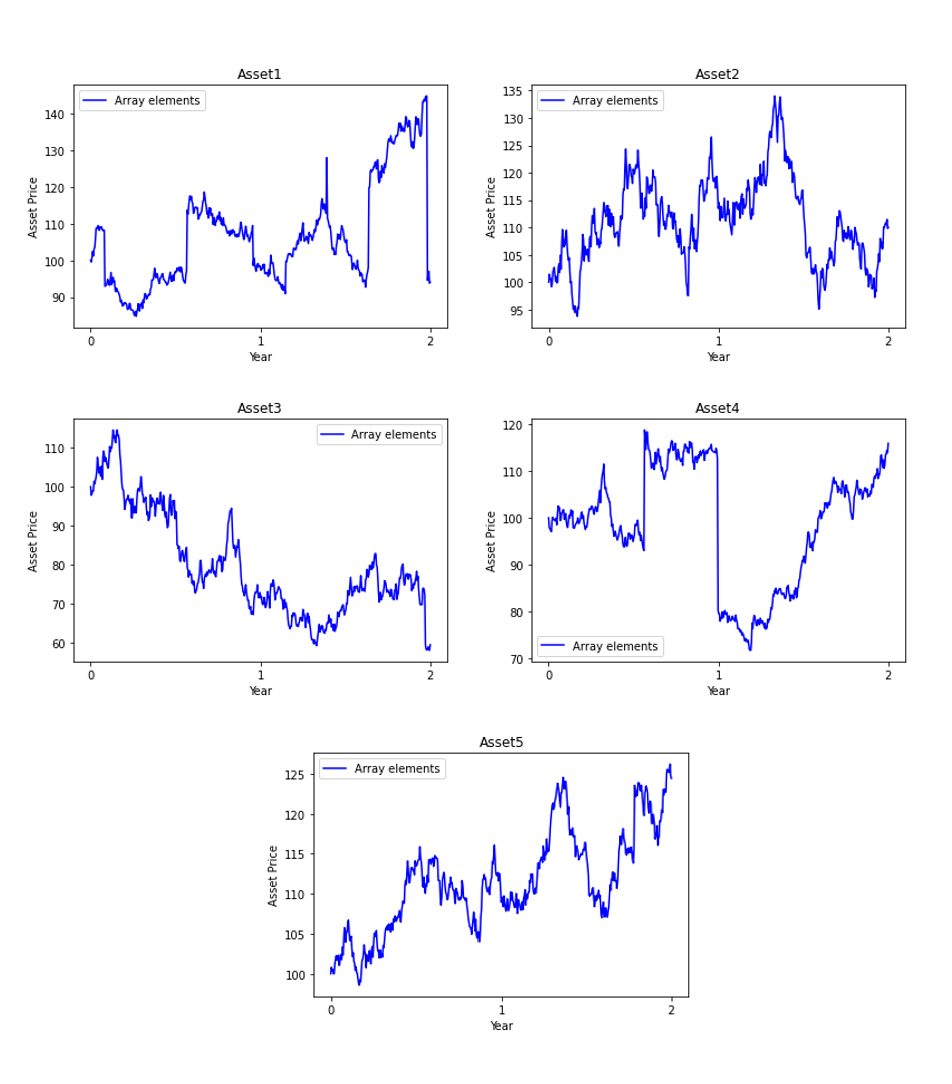}
    \caption{5-asset simulation with risk neutral}
    \label{fig:5assets_rn}
\end{figure}

\section{Hedging Problem}

\subsection{Description of the Hedging Problem} \label{sec:hedging}
A contingent claim paying $g(S_T)$ at time $T$ is applied in this problem. $S_T$ denotes the d-dimension contingent claim underlying vector while function g denotes a mapping transformation from d dimension to 1 dimension. Volume risk introduced in section 2 is the first source of incompleteness. A finite set of hedging dates $t_0<t_1<...<t_{N_1}<t_N=T$(T=2 years, N=504 days) is set as the second source of market incompleteness.  Trades of each of the risky assets $F_i$ are limited with liquidity as a finite quantity $l_i$, which is the third source of incompleteness.A self-financing portfolio is a d-dimensional $(\mathcal{F}_t)$-adapted process $\Delta_t$.$p_i$ denotes initial investment value on asset i. Terminal value $X^{\Delta}_T$ of this model at time T satisfies:
\begin{equation}
    X^{\Delta}_T=p+\sum^d_{i=1}\sum^{N_1}_{j=0}\Delta^i_{t_j}(F^i_{t_{j+1}}-F^i_{t_j})
\end{equation}
The change in one time interval is $C_t^i=\Delta^i_{t_j}-\Delta^i_{t_{j-1}}$ which cannot exceed trading liquidity limits $l_i$:
\begin{equation}
    |\Delta^i_{t_0}|{\leq}l^i,|C^i_j|{\leq}l^i,j=1,...,N-1,i=1,...,d.
\end{equation}
we are searching for a optimized strategy for this hedging problem satisfying:
\begin{equation}
    (p^{Opt},\Delta^{Opt})=Argmin_{p,\Delta}L(X^{\Delta}_T-g(S_T))=Argmin_{p,\Delta}L(Y_T)
\end{equation}
\begin{equation}
    Y_T=X^{\Delta}_T-g(S_T)
\end{equation}
\subsection{Loss Function} \label{sec:parameter-estimation}
Two types of loss functions will be minimized in our research:
\begin{itemize}
\item[*] \textbf{Mean Squared error} 
    \begin{equation}
        L(Y)=\mathbb{E}[Y^2]
    \end{equation}
    It has been studied in Schweizer (1999) with the drawback of penalizing losses and gain the same way.
\item[*] \textbf{Huber Loss Function}
    \begin{equation}
L_{\delta}(y, f(x))=\left\{\begin{array}{ll}
\frac{1}{2}(y-f(x))^{2}, & |y-f(x)| \leq \delta \\
\delta|y-f(x)|-\frac{1}{2} \delta^{2}, & |y-f(x)|>\delta
\end{array}\right.
    \end{equation}
    The Huber loss function involves a hyper-parameter $\delta$. It has quadratic loss for $|Y|\leq\delta$ and linear loss for $|Y|\ge\delta$, so it is more robust to outliers compared to the MSE Loss.
\end{itemize}

\section{Neural Networks}
Three well-studied neural networks  is proposed in Gao(2021) to solve the hedging problem in one dimension: Recurrent Neural Network(RNN); Long-Short term Memory(LSTM) which solve the vanishing gradient problem of RNN; Mogrifier-LSTM which is first put forward by Melis et al.(2020) introducing "mutual gates" between hidden layer and input state. It shown that loss in the three algorithms all converged fast while in terms of the non-risk-neutral data, more jumps were observed. In this paper, we adapt the three algorithms to our model and study whether they also achieve good results in one dimension, then we use the best-performing algorithm to solve the five-dimension optimization problem.
\subsection{RNN and LSTM}
RNN is a recursive neural network which recurses in sequence evolution direction.In RNN, the imput is time series data(namely $S_t$) which are imported into the network successfully. A memory state $h_t$ and an output state $C_t$. At each time step t, a recurrent cell in a hidden state connect the input $x_t$ with the previous information $h_{t-1}$ and derive the output of $h_t$ and $C_t$ which continues the iteration. From the vertical perspective, $h_t$ is transformed from the front layer as an input of $x_t$ at the previous layer.
LSTM cells are capable of dealing with long-range data and solve the vanishing gradient problem where RNN falls short.
Gates in an LSTM cell consist of three parts: forget gate$\mathcal{F}_t$,input gate$\mathcal{I}_t$,output gate $\mathcal{O}_t$. The gates control the information flow, y adding or moving information in the memory state $h_t$. The explicit operation of the gates follows:
\begin{equation}
\begin{array}{r}
\mathcal{F}_{t}=\sigma\left(Q_{F} S_{t}+R_{F} C_{t-1}+b_{F}\right) \\
\mathcal{I}_{t}=\sigma\left(Q_{I} S_{t}+R_{I} C_{t-1}+b_{I}\right) \\
\mathcal{O}_{t}=\sigma\left(Q_{O} S_{t}+R_{O} C_{t-1}+b_{O}\right) \\
h_{t}=\mathcal{F}_{t} \odot h_{t-1}+\mathcal{I}_{t} \odot \tanh \left(Q_{h} S_{t}+R_{h} C_{t-1}+b_{h}\right), h_{0}=0 \\
C_{t}=\mathcal{O}_{t} \odot \tanh \left(h_{t}\right), C_{0}=0
\end{array}
\end{equation}

where $\odot$ is the Hadamard product, $\sigma$ is the sigmoid activation function ($\sigma \left( x\right) =\frac{1}{1-e^{-x}}$),$Q_{\star}\in \mathbb{R}^{h\times d}$,$R_{\star}^{h\times h}$,$b_{\star}\in \mathbb{R}^h$
,h representing cell state size. The forget gate $\mathcal{F}_{t}$ decides which information should be removed from the memory state.The input gate $\mathcal{I}_t$ processed the new information and the output gate $\mathcal{O}_t$ evaluate the memory state and decide the outputtings. Throughout the procedure, the weight matrices and bias vector ($Q_{\star}$,$R_{\star}$,$b_{\star}$) keep unchanged at each time step.$C_t$ is outputted as an approximation of the unknown function.

\begin{figure}
    \centering
    \includegraphics[scale=0.37]{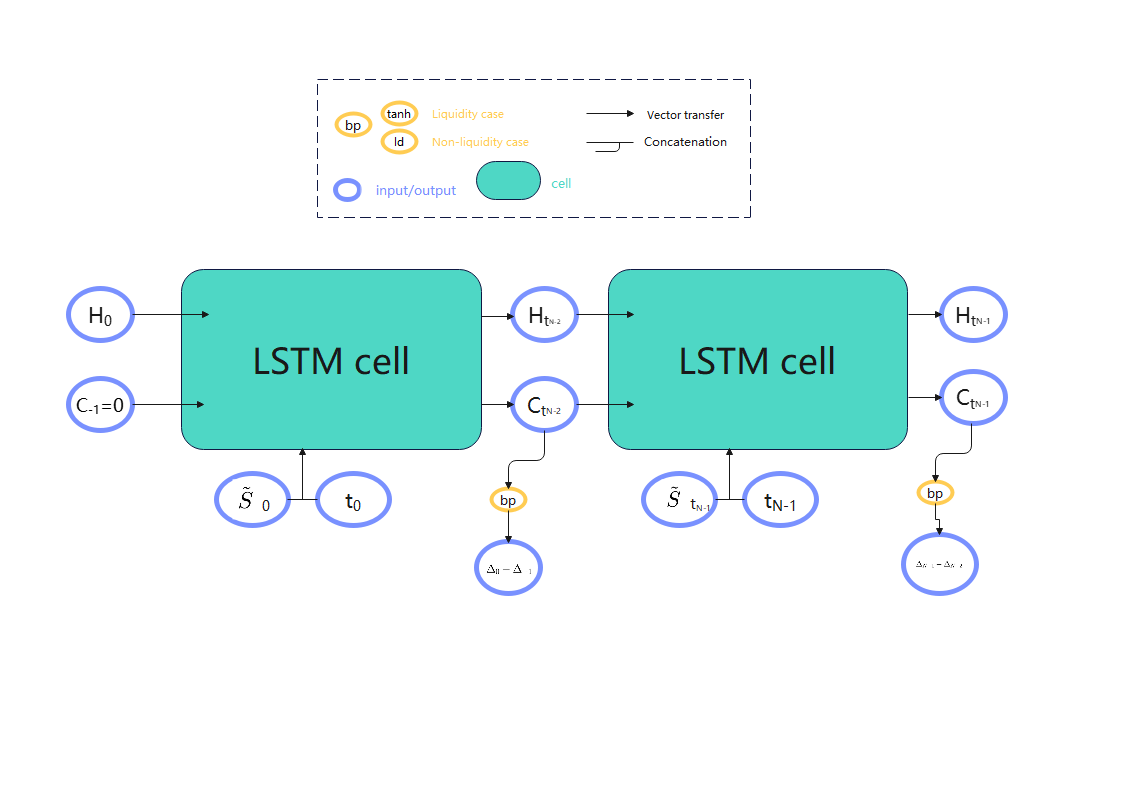}
    \caption{LSTM cell structure}
    \label{fig:LSTM cell structure}
\end{figure}

\subsection{Mogifier LSTM}
Based on classic LSTM, Mogrifier-LSTM introduces "mutual gating" of the current input $x_t$ and memory state $h_t$ which enable Mogrifier-LSTM performs better in natural language processing problems.

The specific equations and diagram for Mogrifier-LSTM follows:
\begin{equation}
    x^i=2\sigma(Z^i h^{i-1}_{prev})\odot x^{i-2}, i\in[1,2,...r],i=2n+1,n\in\mathbb{Z}
\end{equation}
\begin{equation}
    h^{i-1}_{prev}=2\sigma(H^i x^{i-1})\odot h^{i-2}_{prev}, i\in[1,2,...r],i=2n,n\in \mathbb{Z}
\end{equation}

\begin{figure}
    \centering
    \includegraphics[scale=1]{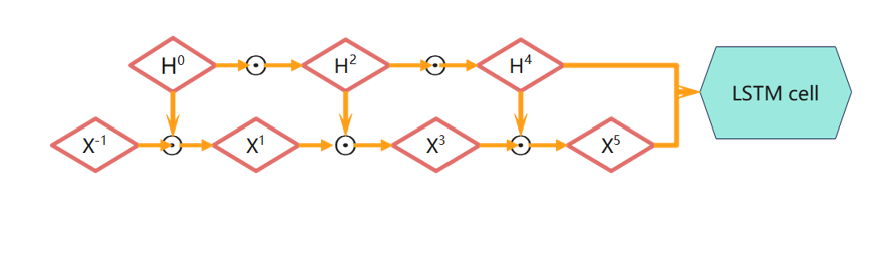}
    \caption{Mogrifier steps}
    \label{fig:mogstep}
\end{figure}

The input at time $t(x_t)$ is $x^{-1}$.$Z^i,H^i$ are the adaptive matrices. r is a hyper-parameter initialized \textbf{4}. The illustration of the whole Mogrifier LSTM cell is as follows:

\begin{figure}
    \centering
    \includegraphics[scale=0.2]{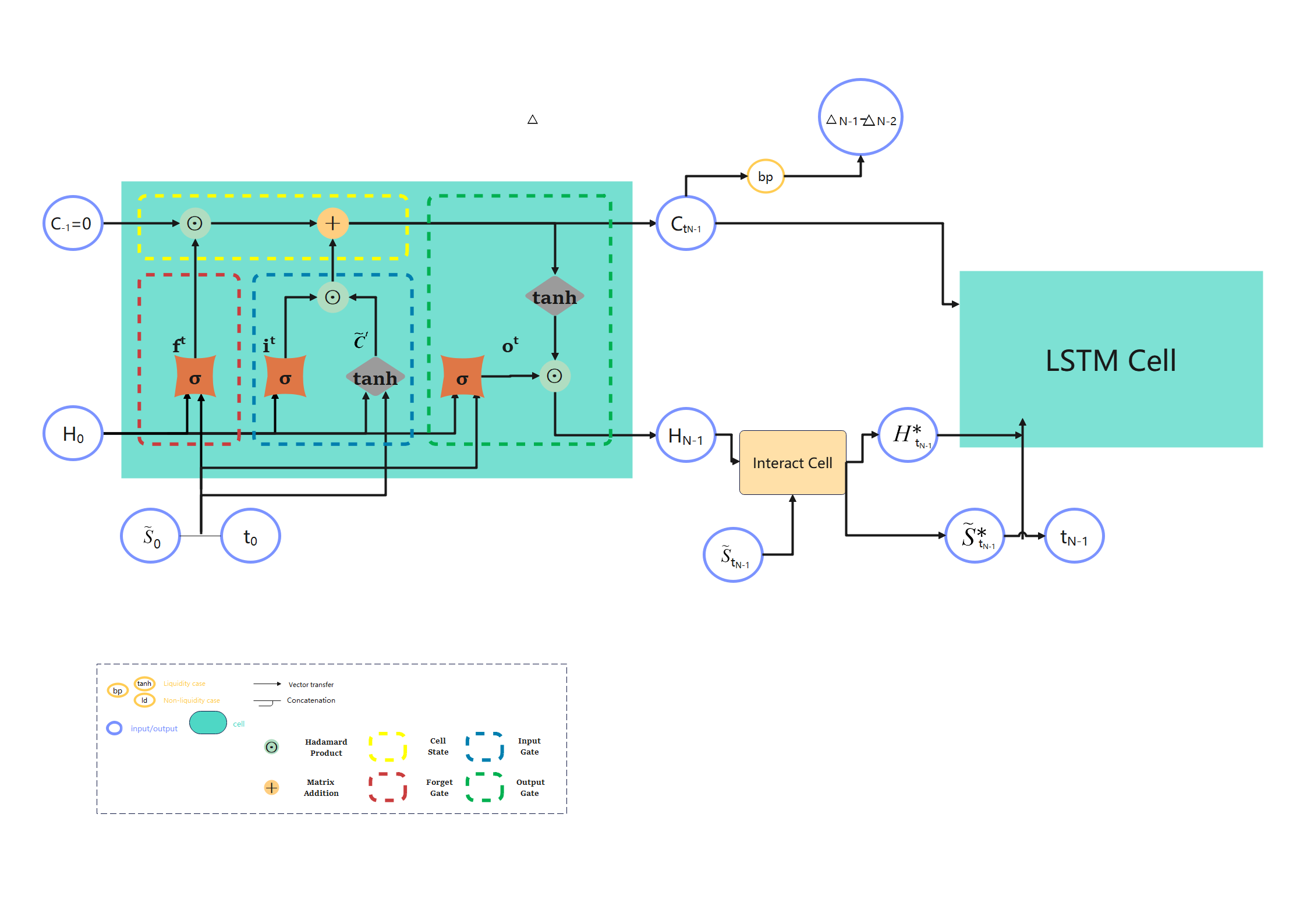}
    \caption{Mogrifier steps}
    \label{Mogrifier steps}
\end{figure}

\subsection{General Optimization Algorithm}
Adaptive Moment Estimation(Adam) \cite{2014Empirical} method is taken to calculate the  parameters $\theta$. It stores an exponentially decaying average of the gradients and keeps the average similar to momentum. 

\subsection{Problem Solving}
A normalized version of $S_t$ which is denoted by $\tilde{S}_t=\frac{S_t-\mathbb{E}[S_t]}{\sqrt{\mathbb{E}[{(S_t-\mathbb{E}(S_t))}^2]}}$ is fed to the three neural networks. According to the hedging problem in Section 3, the final payoff can be specified as:
\begin{equation}
     X_T(\theta)=p(\theta)+\sum^d_{i=1}\sum^{N_1}_{j=0}\Delta^i(t_j,(\tilde{S}_{t_{k}})_{k\leq j},\theta)(F^i_{t_{j+1}}-F^i_{t_j}).
\end{equation}
where $p(\theta)$ denotes the premium and $\Delta_{t_{j}}$ the transactions at date $t_{j}$.
As is shown in Fecamp et al(2020), the recurrent cell at the time step $t_j$ is fed with $\tilde{S}_{t_{j}}$ and output a d-dimensional matrix $\hat{C}_j(\theta,(\tilde{S}_{t_{s}})_{s\leq j},(\Delta_{t_s})_{s\leq j})$.The strategy is denoted by $\Delta^i$, $i=1,...,d$
\begin{equation}
    \Delta^i(t_j,(\tilde{S}_{t_{i}})_{i\leq j},\theta)=\sum_{k=0}^{j} \hat{C}_k^i((\tilde{S}_{t_{s}})_{s\leq j}, \Delta(t_s,(\tilde{S}_{t_{s}})_{s\leq j},\theta)).
\end{equation}
and the optimization problem becomes:
\begin{equation}
    \theta^{*}=Argmin_{\theta}L(X_T(\theta)-g(S_T)).
\end{equation}

The incompleteness of illiquidity is described in Section 3.1.According to Fecamp(2021),the buy and sell order is $C$. When operating the neural network, the output is specified as follows:
\begin{equation}
    \Delta^i(t_j,(\tilde{S}_{t_{i}})_{i\leq j},\theta)=l^i\sum_{k=0}^{j}tanh(\hat{C}_k^i((\tilde{S}_{t_{s}})_{s\leq j}, \Delta(t_s,(\tilde{S}_{t_{s}})_{s\leq j},\theta))).
\end{equation}
Apparently, all $\Delta$'s involved are within the interval$[-l_i,l_i]$.

For one-dimensional problem:
The three output layers are: \textbf{Premium layer} with output size 1,\textbf{Transaction layer} with output size $N$(all elements within the range [-$l$,$l$]),and \textbf{Transaction date layer} with output size $T$. To show the probability of a transaction in each day, sigmoid transformed the result of transaction date layer into a probability space and we take the N dates with the highest probability.

For d-dimensional problem:
We have two output layers.\textbf{Premium layer} has an output size of 1.For each time step $t_j,$\textbf{Transaction layer} has an output size of $D$(all elements within the range [-$l_i$,$l_i$]).Different from one-dimensional problem, we choose N trading dates instead of using transaction date layer to choose the N transaction dates with the highest probability.

\subsection{Hyper-parameters}
The hyper-parameters in neural networks for one-dimentional problem are listed as follows:

\begin{itemize}
    \item[$\bullet$] The batch size is 20.
    
    \item[$\bullet$] Adam optimizer is adapted in all networks with initial learning rate $10^{-3}$.
    
    \item[$\bullet$] Each hidden layer has 50 LSTM units.
    
    \item[$\bullet$] Epoch number is set to be $1000$.
    
    \item[$\bullet$] $\delta=0.5$ in the Huber loss function.
    
    \item[$\bullet$] $r=5$ for the Mogrifier-LSTM.
    
\end{itemize}
For multi-dimensional problem, the batch size is changed into 50 and the epoch number of loss iteration with risk neutral is 253 (because of the extreme operation time) and the epoch number of loss iteration without risk neutral is $1000$.

\section{Results For Hedging Problem}
In this part we first compare the performance of the three neural networks on the two loss function in 3.2. Next, five assets are involved to make a illustration on multi-dimension hedging problem.

\subsection{Comparison in one-dimension problem}\label{sec:comparsion}
Numerical results of loss functions in this hedging problem optimized by three neural networks (RNN,LSTM,Mogrifier-LSTM) are plotted in the following figures with 1000 epochs.Pictures on the right refer to the performance in the last 20 epochs for sake of final loss observation.
Figure1-2 and 3-4 in the following pages demonstrate the loss of two JD-models 2 accordingly. 

\begin{figure}
    \centering
    \includegraphics[scale=1.05]{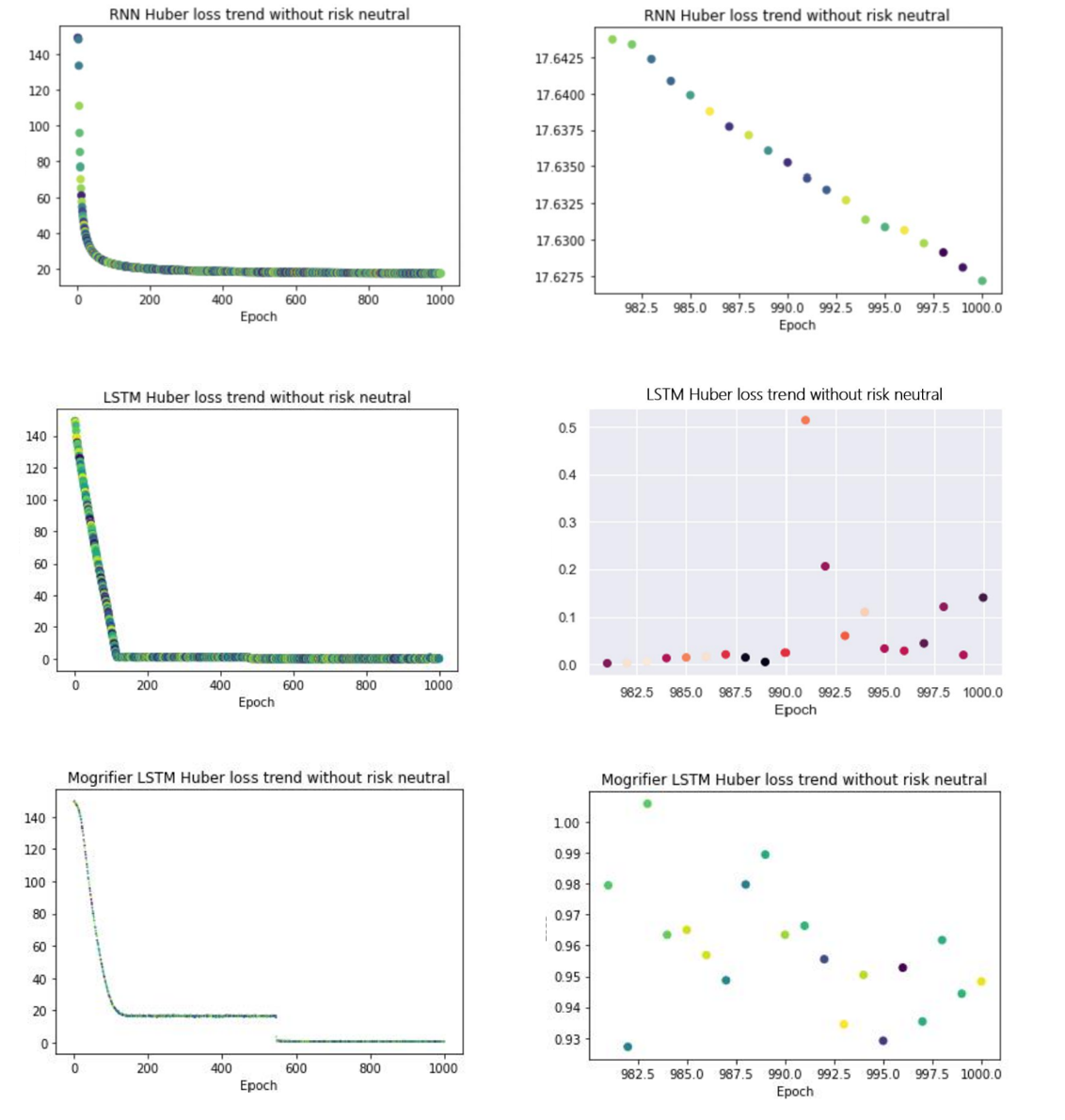}
    \caption{Huber loss trend without risk neutral}
    \label{fig:HB no risk neutral}
\end{figure}
\begin{figure}
    \centering
    \includegraphics[scale=1.05]{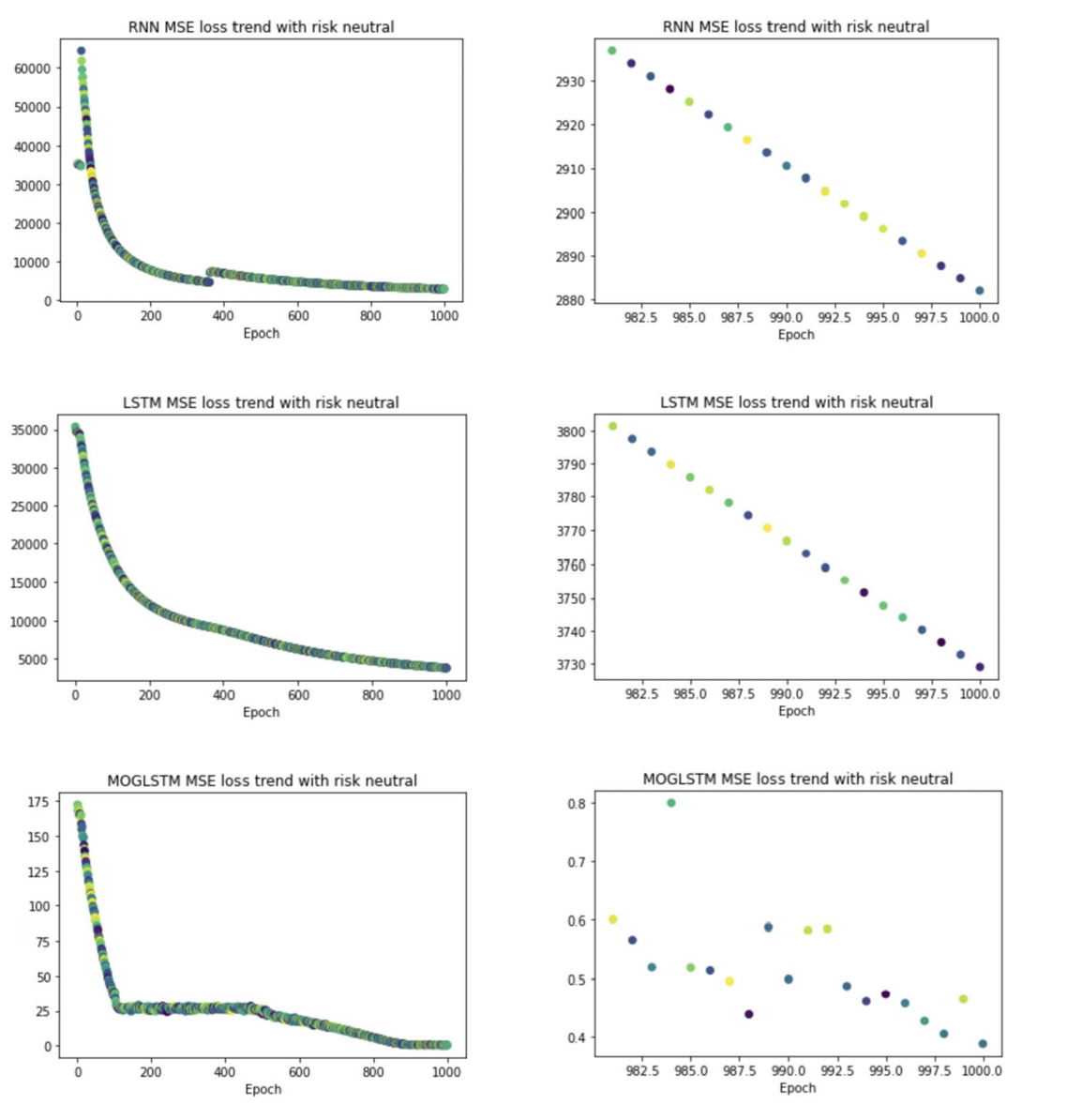}
    \caption{MSE loss trend with risk neutral}
    \label{fig:MSE rna}
\end{figure}
\begin{figure}
    \centering
    \includegraphics[scale=1.05]{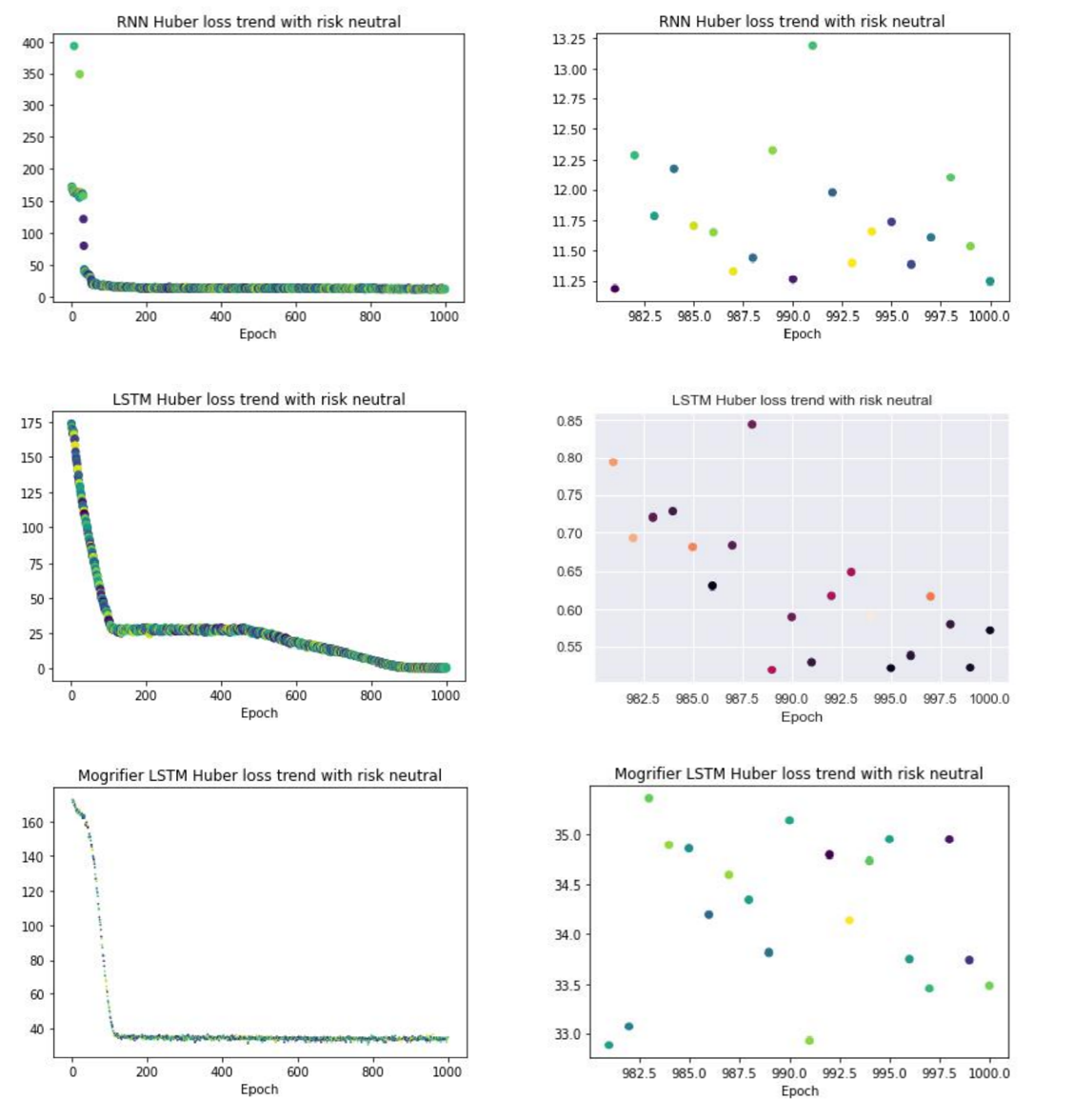}
    \caption{Huber loss trend with risk neutral}
    \label{fig:MSE rn}
\end{figure}

All models have good performances in this problem. However, it's apparent that Mogrifier-LSTM loss has faster convergence whether with or without risk neutral. Besides, it's obvious that Mogrifier-LSTM MSE loss seems to be steady in the last 20 epoches while the other two are still decreasing. Also, appearance of Huber Loss is generally better than MSE Loss, which corresponds to robustness of Huber Loss. Loss values without and with risk neutral are respectively listed in Table 2 and 3.

\vspace{6mm}
\begin{table}
    \centering
    \begin{tabular}{c|cc}
                       & \textbf{MSE Loss} & \textbf{Huber loss} \\ \hline
\textbf{RNN}           & 7.51E+02       &  1.76E+01              \\ \hline
\textbf{LSTM}           & 9.01E+02      &  1.80E-02             \\ \hline
\textbf{Mogrifier-LSTM} & 2.03E-02      &  9.48E-01             \\
        \end{tabular}
        \caption{Neural Networks Loss Without Risk Neutral}
        \label{tab:store-size}
\vspace{4mm}
\end{table}
\begin{table}
    \centering
    \begin{tabular}{c|cc}
                       & \textbf{MSE Loss} & \textbf{Huber loss} \\ \hline
\textbf{RNN}           & 2.88E+03       &  1.13E+01              \\ \hline
\textbf{LSTM}           & 3.73E+03      &  5.81E-01             \\ \hline
\textbf{Mogrifier-LSTM} & 3.90E-01      &  3.35E+01             \\
        \end{tabular}
        \caption{Neural Networks Loss With Risk Neutral}
        \label{tab:store-sizeb}
\vspace{4mm}
\end{table}

The abnormal values of loss function in front charts are probably the result of overfitting, especially in one dimension. The impact of risk neutral is unconspicuous in this hedging problem  so reasons for this phenomenon need to be further studied.

\subsection{Analysis of Mogrifier-LSTM}\label{sec.analysis}

\begin{figure}
    \centering
    \includegraphics[scale=0.4]{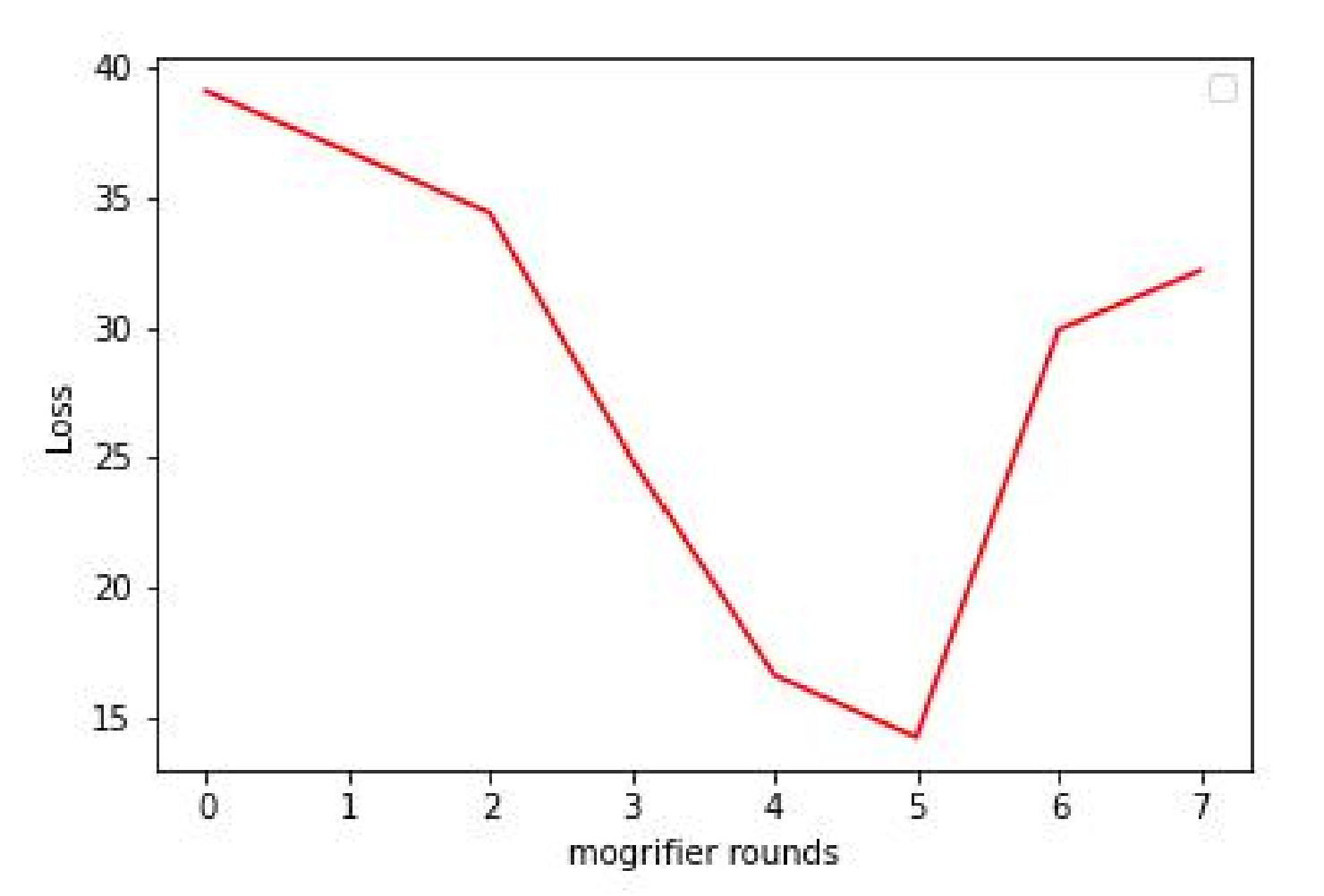}
    \caption{Loss in different Mogrifier rounds}
    \label{fig:loss in rounds}
\end{figure}

\begin{figure}
    \centering
    \includegraphics[scale=0.7]{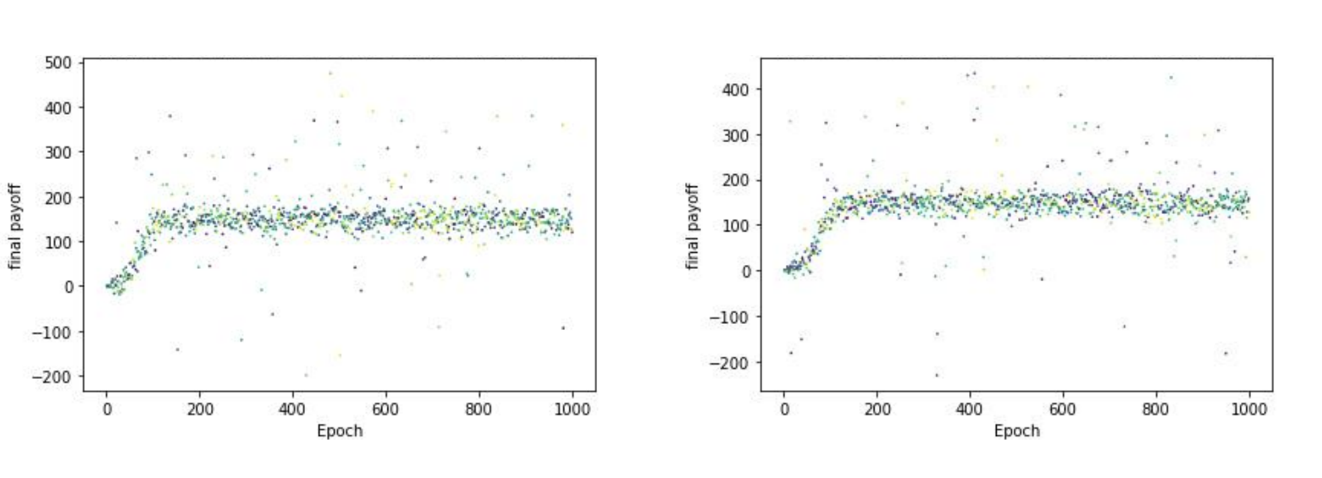}
    \caption{Payoff trend of Mogrifier-LSTM in MSE and Huber loss}
    \label{fig:loss in roundsa}
\end{figure}
In Table 4, Mogrifier-LSTM Loss values with different mogrifier rounds are calculated. $L_i$ denotes the loss value at time $t$ in round $i$. In each round epochs are set as 1000. \\
\begin{table}
    \centering
    \begin{tabular}{c|c}
                        & \textbf{Huber loss} \\ \hline
\textbf{$0_th$ Round}               &  3.91E+01              \\ \hline
\textbf{$1_st$ Round}               &  3.68E+01             \\ \hline
\textbf{$2_nd$ Round}               &  3.44E+01             \\ \hline
\textbf{$3_rd$ Round}               &  2.49E+01             \\ \hline
\textbf{$4_th$ Round}               &  1.66E+01              \\ \hline
\textbf{$5_th$ Round}               &  1.42E+01              \\ \hline
\textbf{$6_th$ Round}               &  2.99E+01              \\ \hline
\textbf{$7_th$ Round}               &  3.22E+01              \\ \hline
        \end{tabular}
        \caption{Loss Values In Different Rounds}
        \label{tab:store-sizec}
\vspace{4mm}
\end{table}
\\It's obvious that results are better when r equals to 5. Furthermore, more rounds only increase computation complexity but don't decrease Huber Loss value.\\
\subsection{Extension to Five Dimensions}
Based on Sec 5.1 and 5.2, we choose Mogrifier-LSTM under the estimation of Huber loss to carry out the 5-dimensional hedging problem in the two jump-diffusion models with the mogrifier round number of 5 because their good performance in dealing with one-dimension hedging problem.

\begin{figure}
    \centering
    \includegraphics[scale=1.05]{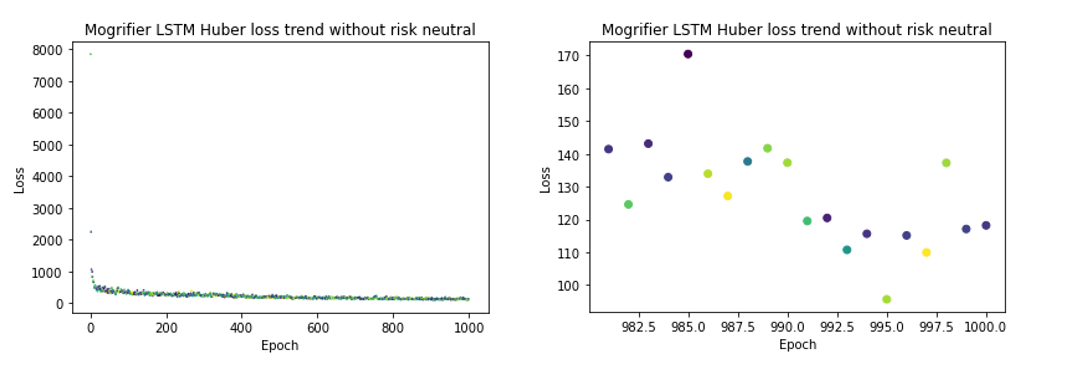}
    \caption{5 assets Huber loss trend of Mogrifier-LSTM without risk neutral}
    \label{fig:5assetloss_wtrn}
\end{figure}
\begin{figure}
    \centering
    \includegraphics[scale=1.05]{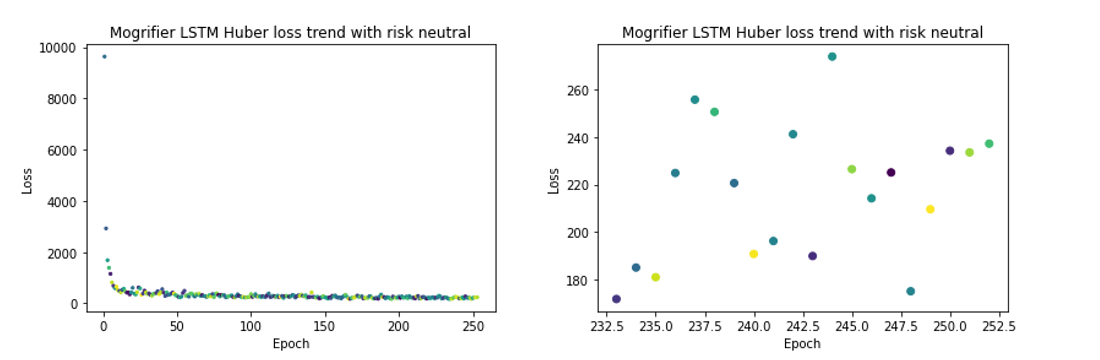}
    \caption{5 assets Huber loss trend of Mogrifier-LSTM with risk neutral}
    \label{fig:5assetloss_rn}
\end{figure}

\begin{table}[H]
\centering
\begin{tabular}{@{}c|ccc@{}}
                       & \textbf{Without Risk-Neutral} & \textbf{With Risk-Neutral} \\ \hline
\textbf{Loss }   &$1.18E+02$($1000_{th}$iteration)        &$2.37E+02$($253_{th}$iteration) \\ \hline
        \end{tabular}
        \caption{Loss of two jump-diffusion models}
        \label{tab:5 assets loss}
\vspace{-4mm}
\end{table}
Loss convergence is observed in both models. The results of neural network is presented in Fig.16 and 17 which meet our expectation and proved the effectiveness of solving multi-dimensional hedging problems with Mogrifier-LSTM with the round number of 5. Table 4 shows the loss of two jump-diffusion models. It is obvious that for five-dimensional hedging problem, the loss is considerable. Because of the relatively long operation time(over 20 hours by intel core $i5$ for 253 iterations) and the stagnation of loss trend (around 200), further proof of the practicability and improvements on algorithm still remain to be proceeded.
\section{Conclusion}

Three neural-network-based algorithms applied to the hedging of contingent claim are proposed. RNN, LSTM and Mogrifier LSTM show good performances in convergence. Particulary, mogrifier LSTM has a faster loss convergence speed than RNN and traditional LSTM which means less time cost on solving this problem. Besides, mogrifier round 5 is suggested for lower loss value.
It's obviously indicated that deep learning neural networks help solve the hedging problems a lot especially in incomplete financial
market (non-continuous prices, discrete transaction time and illiquidity).

\bibliographystyle{unsrt}  
\bibliography{references}

\begin{thebibliography}{1}

\bibitem{2008Bilateral}
U.~K. Tappe.
\newblock Bilateral gamma distributions and processes in financial mathematics.
\newblock {\em Stochastic Processes and their Applications}, 2008.

\bibitem{2001Purely}
D.~B. Madan.
\newblock Purely discontinuous asset price processes.
\newblock {\em option pricing interest rates and risk management}, 2001.

\bibitem{fecamp2020deep}
Simon FECAMP, Joseph MIKAEL, and Xavier WARIN.
\newblock Deep learning for discrete-time hedging in incomplete markets.
\newblock {\em Journal of computational Finance}, 2020.

\bibitem{0Inflexible}
Y.~Gao, Y.~Wu, and M.~Duan.
\newblock Inflexible hedging in the presence of illiquidity and jump risks.
\newblock {\em Social Science Electronic Publishing}, 2021.

\bibitem{Matsuda2004IntroductionTM}
Kazuhisa Matsuda.
\newblock Introduction to merton jump diffusion model.
\newblock 2004.

\bibitem{Kou2002AJM}
Steven Kou.
\newblock A jump-diffusion model for option pricing.
\newblock {\em Manag. Sci.}, 48:1086--1101, 2002.

\bibitem{2008Hooke}
Lahouaria Benasla, Abderrahim Belmadani, and Mostefa Rahli.
\newblock Hooke-jeeves.
\newblock {\em Journal of Information Science and Engineering}, 24(3):907--917,
  2008.

\bibitem{2014Empirical}
J.~Chung, C.~Gulcehre, K.~H. Cho, and Y.~Bengio.
\newblock Empirical evaluation of gated recurrent neural networks on sequence
  modeling.
\newblock {\em Eprint Arxiv}, 2014.

\end{thebibliography}

\end{document}